\documentclass[a4paper]{jpconf}
\usepackage{graphicx}
\begin{document}
\title{The nucleon axial mass and the MiniBooNE CCQE neutrino-nucleus
 data}

\author{J Nieves$^1$, I Ruiz Simo$^2$ and M J  Vicente Vacas$^3$}

\address{$^1$ Instituto de F\'\i sica Corpuscular (IFIC), Centro Mixto
Universidad de Valencia-CSIC, Institutos de Investigaci\'on de
Paterna, E-46071 Valencia, Spain}

\address{$^2$  Departamento de
  F\'{\i}sica At\'omica, Molecular y Nuclear,\\
  Universidad de Granada, E-18071 Granada, Spain.}

\address{$^3$ Departamento de F\'\i sica Te\'orica and IFIC, Centro Mixto
Universidad de Valencia-CSIC, Institutos de Investigaci\'on de
Paterna, E-46071 Valencia, Spain}

\ead{jmnieves@ific.uv.es}

\begin{abstract}
 We analyze the MiniBooNE CCQE $d\sigma/dT_\mu d\cos\theta_\mu$ data
 using a theoretical model that has proved to be quite successful in
 the analysis of nuclear reactions with electron, photon and pion
 probes. We find that RPA and multinucleon knockout turn out to be
 essential for the description of the MiniBooNE data. We show these
 measurements are fully compatible with former determinations of
 nucleon axial mass $M_A$, in contrast with several previous analyses,
 which have suggested an anomalously large value. We find, $M_A=1.08
 \pm 0.03$ GeV. We also argue that the procedure, commonly used to
 reconstruct the neutrino energy for QE events from the muon angle and
 energy, could be unreliable for a wide region of the phase space, due
 to the large importance of multinucleon events.
\end{abstract}

{\small \it 
Contribution to NUFACT 11, XIIIth International Workshop on Neutrino
Factories, Super beams and Beta beams, 1-6 August 2011, CERN and
University of Geneva (Submitted to IOP conference series).}

\section{Introduction}

The interaction of neutrinos with nuclei at intermediate energies
provides relevant information on the axial hadronic currents. The
predicted cross sections for charged current (CC) quasielastic (QE)
scattering are very similar for most theoretical models though,
however, they are clearly below the recently published MiniBooNE
data~\cite{AguilarArevalo:2010zc}. Actually, the cross
section per nucleon on $^{12}$C is clearly larger than for free
nucleons. The discrepancy is large enough to
provoke much debate and theoretical attention. Some works try to
understand these new data in terms of a larger value of $M_A$ (nucleon
axial mass) around
1.3--1.4 GeV~\cite{AguilarArevalo:2010zc,
Butkevich:2010cr,Benhar:2010nx,Juszczak:2010ve}. These large values
are not only difficult to accommodate theoretically, but are also in
conflict with the value for $M_A=1.03\pm 0.02$ GeV that is usually
quoted as the world
average~\cite{Bernard:2001rs,Lyubushkin:2008pe}. 

\begin{figure}[h]
\includegraphics[width=25pc]{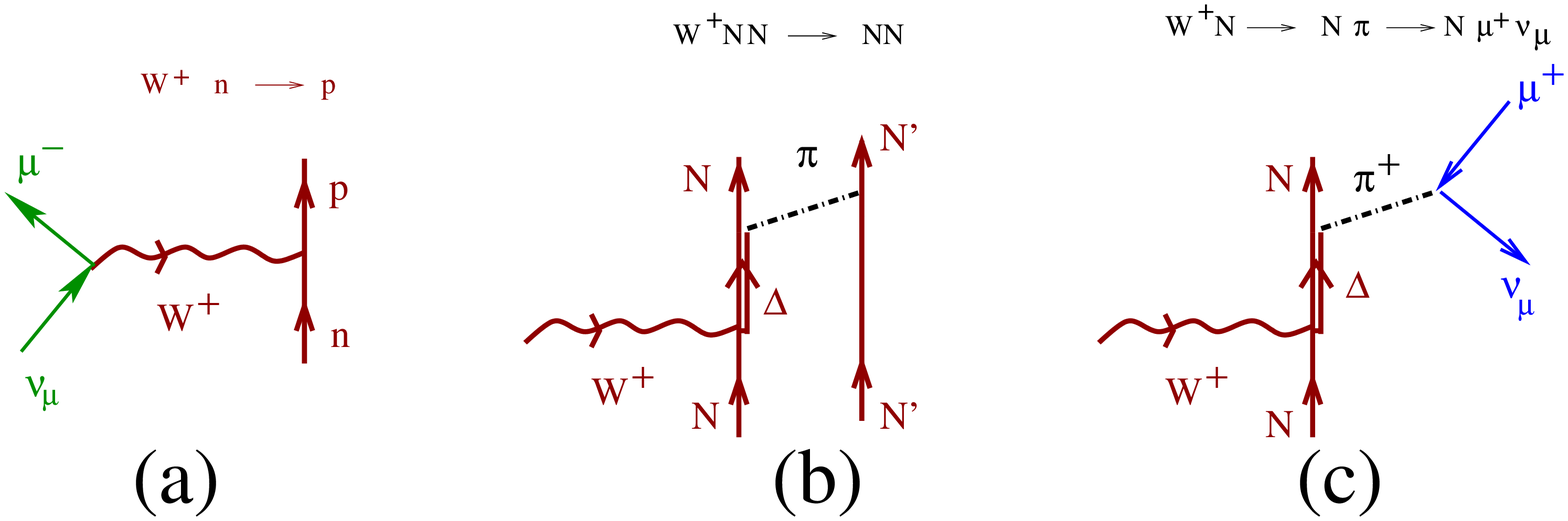}\hspace{2pc}%
\begin{minipage}[b]{11pc}\caption{\label{fig:expl}
Mechanisms for $W$ absorption inside of a nucleus.}
\end{minipage}
\end{figure}

In most theoretical works QE is used for processes where the gauge
boson $W$ is absorbed by just one nucleon, which together with a
lepton is emitted (see Fig.~\ref{fig:expl}a). However, in the recent
MiniBooNE measurements, QE is related to processes in which only a
muon is detected. This latter definition could make sense because
ejected nucleons are not detected in that experiment, but includes
multinucleon processes (see Fig.~\ref{fig:expl}b) and others like pion
production followed by absorption\footnote{Note, MiniBooNE analysis
Monte Carlo corrects for those events.}. However, it discards pions
coming off the nucleus, since they will give rise to additional
leptons after their decay (see Fig.~\ref{fig:expl}c). In any case,
their experimental results cannot be directly compared to most
previous calculations, as it was first pointed out by M. Martini et
al.~\cite{Martini:2009uj,Martini:2010ex}, in which only the one-body
QE contribution is considered.
In this talk, we present a microscopic calculation of the CCQE-like
double differential cross section $\frac{d\sigma}{dT_\mu
d\cos\theta_\mu}$ measured by MiniBooNE and we will use these data to
extract $M_A$.  We will estimate the CCQE-like cross section by the
sum of the theoretical QE ( Fig.~\ref{fig:expl}a) cross section and
that induced by multinucleon mechanisms, as the one depicted in
Fig.~\ref{fig:expl}b, where the gauge boson is being absorbed by two
or more nucleons without producing pions.

\section {MiniBooNE CCQE-like cross sections and
  multinucleon mechanisms}

First we pay attention to  the
total cross section  and compare our results with the 
unfolded data of Ref.~\cite{AguilarArevalo:2010zc}.  Our microscopic
model, derived in Refs.~\cite{Nieves:2011pp} and~\cite{Nieves:2004wx},
starts from a relativistic local Fermi gas (LFG) picture of the
nucleus, which accounts for Pauli blocking and Fermi
motion. The QE contribution was studied in \cite{Nieves:2004wx}
incorporating several nuclear effects. The main one is the medium
polarization (RPA), including $\Delta$-hole degrees of freedom and
explicit $\pi$ and $\rho$ meson exchanges in the vector-isovector
channel of the effective nucleon-nucleon interaction. The model for
multinucleon mechanisms (not properly QE but included in the MiniBooNE
data~\cite{AguilarArevalo:2010zc}) is fully discussed in
Ref.~\cite{Nieves:2011pp}. The model includes one, two, and even
three-nucleon mechanisms, as well as the excitation of $\Delta$
isobars.  There are no free parameters in the description of nuclear
effects, since they were fixed in previous studies of photon,
electron, and pion interactions with
nuclei~\cite{Carrasco:1989vq,Nieves:1993ev,Nieves:1991ye,Gil:1997bm}.
This theoretical model has proved to be quite successful in
the study of nuclear reactions with 
photon~\cite{Carrasco:1989vq}, 
pion~\cite{Nieves:1993ev,Nieves:1991ye} and electron~\cite{Gil:1997bm}
probes. 

\begin{figure}[h]
\begin{minipage}{17pc}
\includegraphics[width=16.15pc]{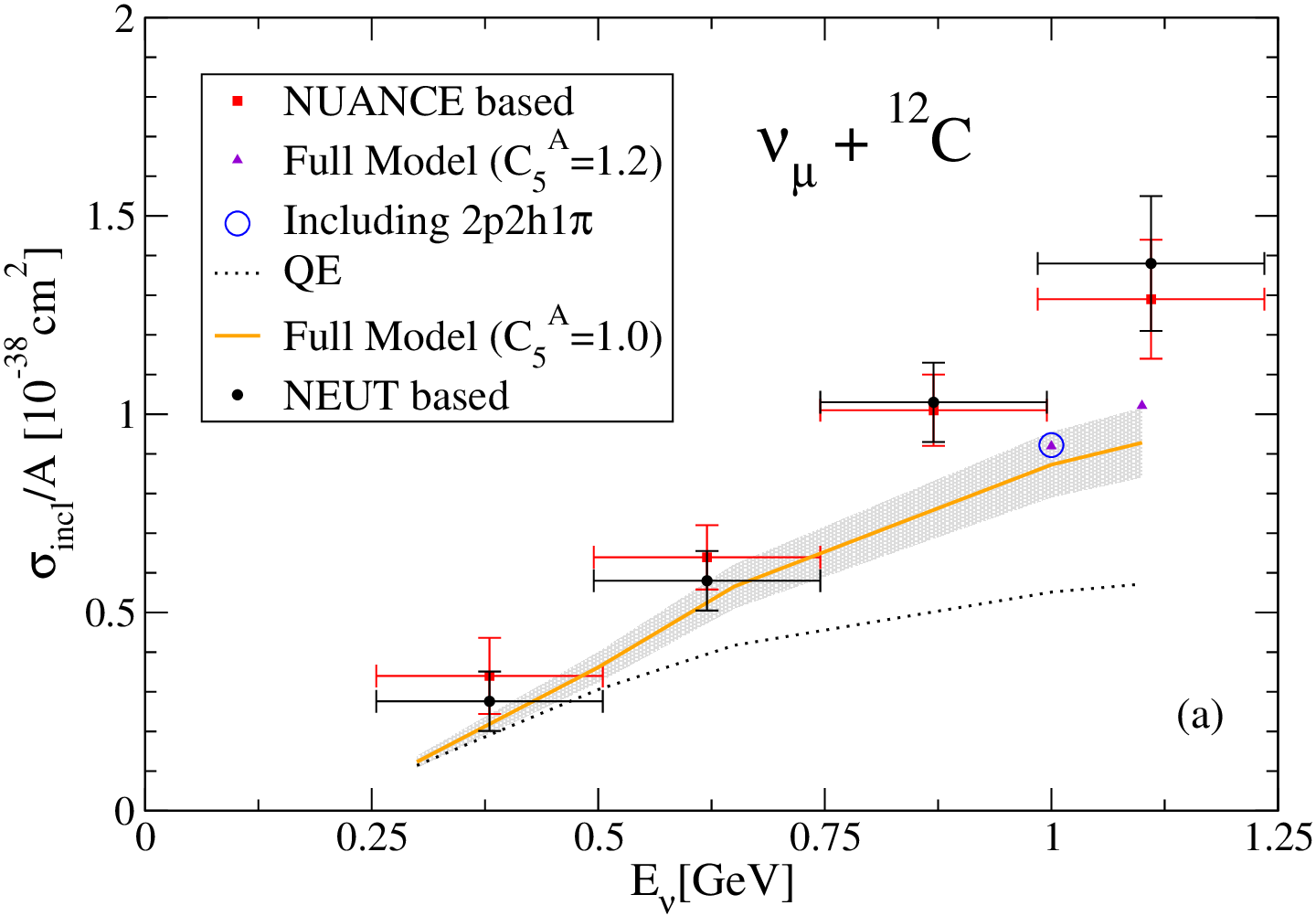}
\caption{\label{fig:incl} SciBooNE neutrino CC inclusive interaction cross
section per nucleon~\cite{Nakajima:2010fp}, together with the QE and
full model (including the theoretical uncertainty band) results of
Ref.~\cite{Nieves:2011pp}.  See this latter reference for further details.
}
\end{minipage}\hspace{3pc}%
\begin{minipage}{17pc}
\includegraphics[width=16.15pc]{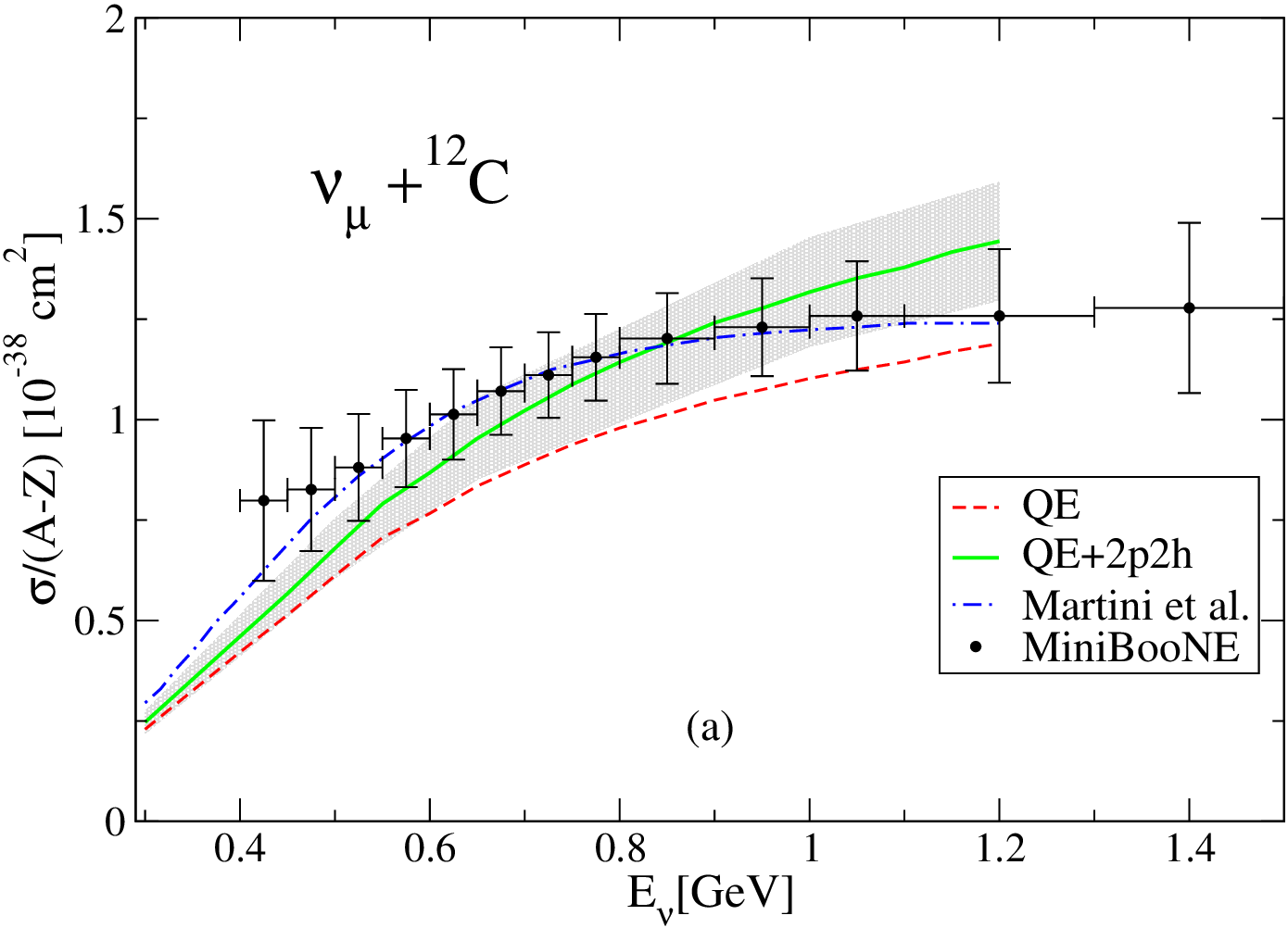}
\caption{\label{fig:mb}Flux-unfolded MiniBooNE $\nu_\mu$ CCQE cross
  section per neutron, together with different theoretical predictions from 
  Ref.~\cite{Nieves:2011pp}. Data points are taken from
  Ref.~\cite{AguilarArevalo:2010zc}. We also show the
  results (blue dash-dotted line) obtained in 
  Ref.~\cite{Martini:2010ex}. 
}
\end{minipage}
\end{figure}

Up to neutrino energies around 1 GeV, the predictions of our model
compare rather well, taking into account experimental and theoretical
uncertainties, with the recent data published by the SciBooNE
collaboration for total neutrino inclusive cross
sections~\cite{Nakajima:2010fp}.  Results are displayed in
Fig.~\ref{fig:incl} taken from Ref.~\cite{Nieves:2011pp}. There, it
can also be appreciated how at larger energies, we underestimate the
cross section. Indeed, we could observe that some $WNN\to NN\pi$
contributions neglected in our model, become relatively important at
these higher energies.

Our predictions~\cite{Nieves:2011pp} for the flux-unfolded muon
neutrino's CCQE-like cross section on $^{12}$C measured
in~\cite{AguilarArevalo:2010zc} are depicted in Fig.~\ref{fig:mb}. The
first observation is that our QE curve misses the data-points, being
our predicted QE cross sections significantly smaller than those
reported by the MiniBooNE collaboration. However, when multinucleon
knock out contributions are added to the QE prediction of
~\cite{Nieves:2004wx}, we obtain the solid green line in a better
agreement with the MiniBooNE data.  In these calculations, $M_A$ is
fixed to 1.05 GeV. Thus we confirm the findings of
\cite{Martini:2009uj,Martini:2010ex} on the crucial role played by the
multinucleon mechanisms in the CCQE-like MiniBooNE data, and that when
these latter processes are considered, high values of $M_A$ in the
1.3-1.4 GeV range are not needed to describe the data. Our evaluation
of these pionless multinucleon emission contributions to the cross
section is fully microscopical and it contains terms, which were
either not considered or only approximately taken into account in
\cite{Martini:2009uj}.

\section{Extracting $M_A$ from MiniBooNE data}
\begin{figure}[h]
\begin{minipage}{21pc}
\includegraphics[width=18.25pc]{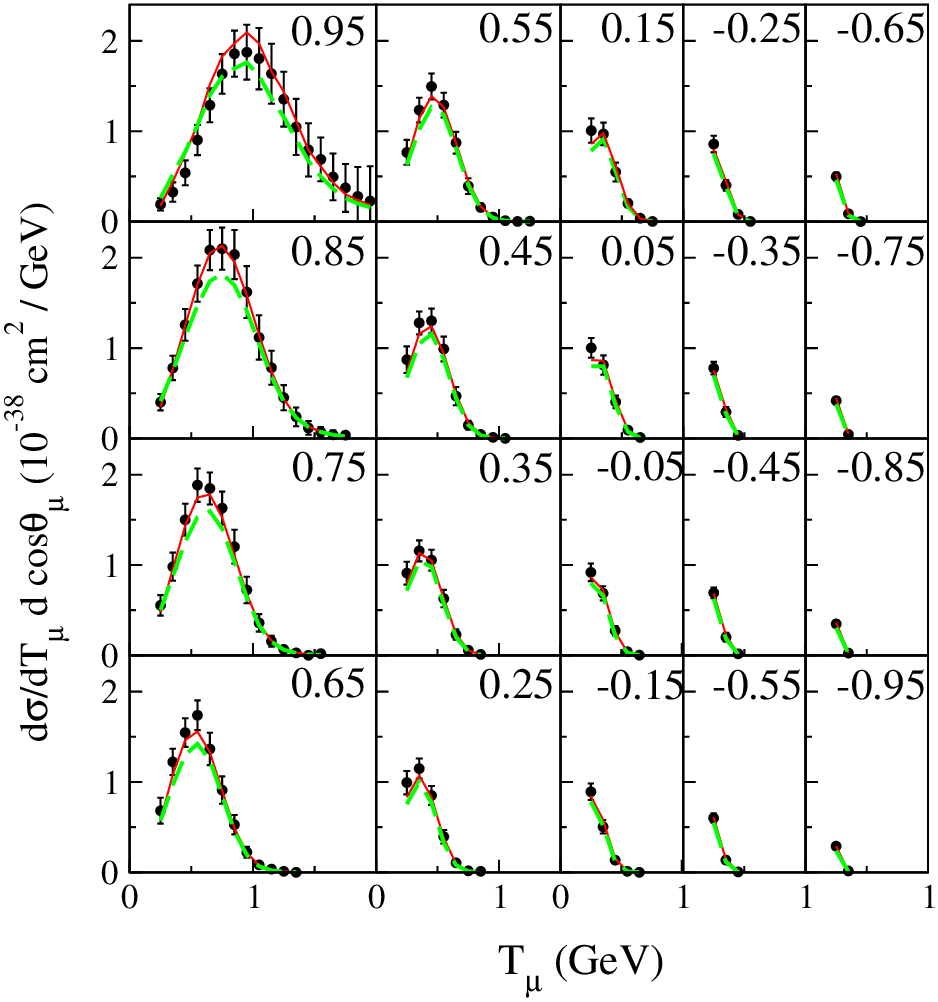}
\caption{\label{fig3}$d\sigma/dT_\mu d(\cos\theta_\mu)$ for different
 angular bins (labeled by its
cosinus central value).  Experimental points are taken from
Ref.~\cite{AguilarArevalo:2010zc}. Green-dashed line (no fit) is the
full model prediction (including multinucleon mechanisms and RPA) of
Ref.~\cite{Nieves:2011pp} and calculated with $M_A=1.049$ GeV.
Red-solid line is best fit ($M_A=1.32$ GeV) for the model without RPA
and without multinucleon mechanisms.}
\end{minipage}\hspace{1.5pc}%
\begin{minipage}{15pc}
\includegraphics[width=15pc]{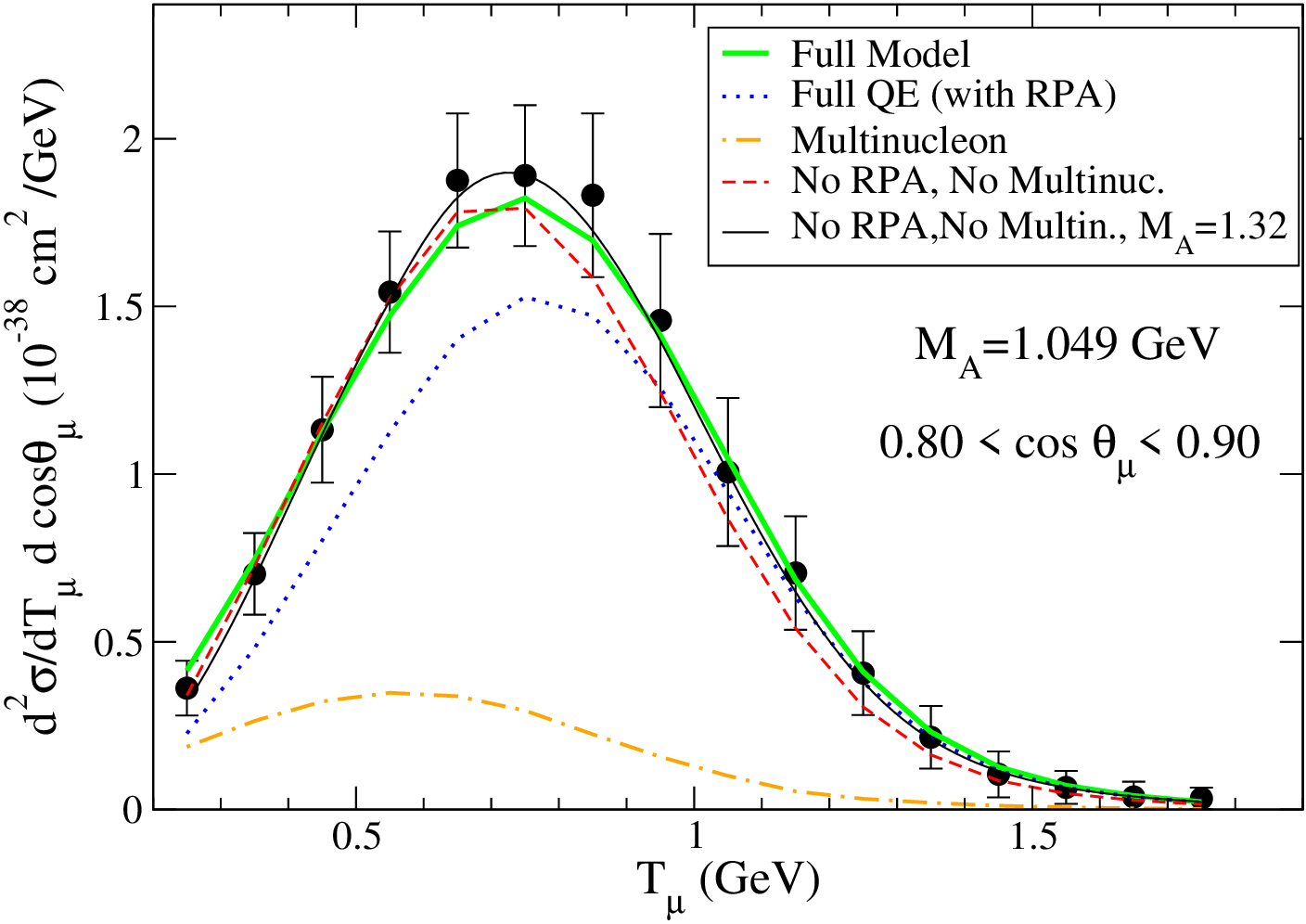}
\caption{\label{fig4}Muon angle and energy distribution for
the $0.80<\cos\theta_\mu<0.90$ bin. Data from
Ref.~\cite{AguilarArevalo:2010zc} and calculation with $M_A=1.32$ GeV
are multiplied by 0.9.  In the other curves a value of $M_A=1.049$
GeV was used.}
\end{minipage} 
\end{figure}

\begin{table}[b]
\caption{\label{tabla}Fit results for various models.}
\begin{center}
\begin{tabular}{llll}
\br
   LFG   & $\lambda=0.96\pm 0.03$   & $M_A =1.32\pm0.03$ GeV & 
$\chi^2$/\# bins=33/137\\ \hline
  Full  & $\lambda=0.92\pm0.03$  & $M_A =1.08\pm0.03$ GeV    &  $\chi^2$/\# bins= 50/137 \\ 
\br
\end{tabular}
\end{center}
\end{table}
The MiniBooNe data include energy and angle distributions as well, and
therefore provide a much richer information. Furthermore, the unfolded
cross section is not a very clean observable after noticing the
importance of multinucleon mechanisms, because the unfolding itself is
model dependent and assumes that the events are purely QE. In
Fig.~\ref{fig3}, we show our results for the MiniBooNE neutrino flux
folded CCQE-like $d\sigma/dT_\mu d\cos\theta_\mu$ distribution with
$M_A=1.049$ GeV (value used in our previous works).  The full
(QE+multinucleon mechanisms) model agrees remarkably well with these
data, despite of that no parameters have been fitted to data, beyond
of a global scale, $\lambda$. The inclusion of this scale $\lambda$
takes into account the global normalization uncertainty of around 10\%
acknowledged in ~\cite{AguilarArevalo:2010zc}.  Details can be found
in \cite{Nieves:2011yp}.  Though the consistency of MiniBooNE data
with standard values of $M_A$ has been established now, one could
still go further and use our full model to fit the data letting $M_A$
to be a free parameter.  We get $M_A=1.08\pm 0.03$ GeV and
$\lambda=0.92\pm0.03$ with a strong correlation between both
parameters. The inclusion of multinucleon mechanisms and RPA is
essential to obtain axial masses consistent with the world
average. This can be appreciated in Fig.~\ref{fig4}, where one can see
that RPA strongly decreases the cross section at low energies, while
multinucleon mechanisms accumulate their contribution at low muon
energies and compensate that depletion.  Therefore, the final picture
is that of a delicate balance between a dominant single nucleon
scattering, corrected by collective effects, and other mechanisms that
involve directly two or more nucleons. Both effects can be mimicked by
using a large $M_A$ value (LFG entry in Table~\ref{tabla}).  We also
see that the proportion of multinucleon events contributing to the
QE-like signal is quite large for low muon energies and thus, the
algorithm used to reconstruct the neutrino energy is badly suited for
this region. This could have consequences in the determination of the
oscillation parameters.
\ack 
This research was supported by DGI and FEDER funds, under
contracts FIS2008-01143/FIS, FIS2006-03438 and CSD2007-00042, by Generalitat
Valenciana contract PROMETEO/2009/0090 and by the EU HadronPhysics2
project, grant agreement n. 227431.  


\section*{References}
\begin{thebibliography}{9}

\bibitem{AguilarArevalo:2010zc}
Aguilar-Areval A A {\it et al.}  [MiniBooNE Collaboration] 2010 {\it
 Phys. Rev.} D {\bf 81}  092005.

\bibitem{Butkevich:2010cr}
  Butkevich A V 2010 {\it Phys. Rev.}  C {\bf 82} 055501.

\bibitem{Benhar:2010nx}
Benhar O, Coletti P and  Meloni D 2010, {\it Phys. Rev. Lett.} {\bf
  105} 132301.

\bibitem{Juszczak:2010ve}
  Juszczak C, Sobczyk J T and Zmuda J 2010
  {\it Phys. Rev. C} {\bf 82} 045502.

\bibitem{Bernard:2001rs}
  Bernard V, Elouadrhiri L and Meissner U G 2002
  {\it J. Phys.} G {\bf 28}.

\bibitem{Lyubushkin:2008pe}
  Lyubushkin V {\it et al.}  [NOMAD Collaboration] 2009
  {\it Eur. Phys. J.  C} {\bf 63} 355.
\bibitem{Nieves:2011pp}  Nieves J, Ruiz Simo I and  and 
Vicente Vacas M J 2011 {\it
 Phys. Rev.} C {\bf 83}  045501.

\bibitem{Nieves:2004wx}
  Nieves J, Amaro J E and Valverde M 2004  {\it
 Phys. Rev.} C {\bf 70}  055503.

\bibitem{Martini:2009uj}
 Martini M, Ericson M,  Chanfray G and 
Marteau J 2009 {\it Phys. Rev.} C {\bf 80}  065501.

\bibitem{Martini:2010ex} Martini M, Ericson M,  Chanfray G and 
Marteau J 2010 {\it Phys. Rev.} C {\bf 81}  045502.


\bibitem{Carrasco:1989vq}
  Carrasco R C  and Oset E 1992
  {\it Nucl. Phys.}  A {\bf 536} 445.

\bibitem{Nieves:1993ev}
  Nieves J, Oset E and Garcia Recio C 1993 
{\it Nucl. Phys.  A} {\bf 554} 509.
\bibitem{Nieves:1991ye}
  Nieves J, Oset E and Garcia Recio C 1993 
{\it Nucl. Phys.  A} {\bf 554} 554.
\bibitem{Gil:1997bm}
  Gil A, Nieves J and Oset E 1997
  {\it Nucl. Phys.}  A {\bf 627} 543.
\bibitem{Nakajima:2010fp}
Nakajima Y {\it et al.}  [SciBooNE Collaboration] 2011 {\it
 Phys. Rev.} D {\bf 83}  012  
\bibitem{Nieves:2011yp} Nieves J, Ruiz Simo I and  and Vicente Vacas M
  J 2011 {\it Preprint} hep-ph/1106.5374

\end{thebibliography}
\end{document}